\documentclass[sigconf]{acmart}
\usepackage[nolist,nohyperlinks]{acronym}
\usepackage{multirow}
\usepackage{bbm}
\usepackage{amsmath}
\usepackage{subfigure}
\AtBeginDocument{%
  }

\usepackage{multirow}
\usepackage{makecell}
\usepackage{graphicx}

\copyrightyear{2024}
\acmYear{2024}
\setcopyright{acmlicensed}\acmConference[RecSys '24]{18th ACM Conference on Recommender Systems}{October 14--18, 2024}{Bari, Italy}
\acmBooktitle{18th ACM Conference on Recommender Systems (RecSys '24), October 14--18, 2024, Bari, Italy}
\acmDOI{10.1145/3640457.3688065}
\acmISBN{979-8-4007-0505-2/24/10}

\setcitestyle{nocompress}

\begin{document}



\title{Do Recommender Systems Promote Local Music? A~Reproducibility Study Using Music Streaming Data}
\author{Kristina Matrosova}
\authornotemark[1]
\affiliation{%
  \institution{CNRS, Geographie-Cités}
  \country{France}
}
\affiliation{%
  \institution{LIPN, USPN}
  \country{France}
}

\author{Lilian Marey}
\affiliation{%
  \institution{LTCI, Télécom Paris}
  \country{France}
}
\affiliation{%
  \institution{Deezer Research}
  \country{France}
}

\author{Guillaume Salha-Galvan}
\affiliation{%
  \institution{Deezer Research}
  \country{France}
}

\author{Thomas Louail}
\affiliation{%
  \institution{CNRS, Geographie-Cités}
  \country{France}
}
\affiliation{%
  \institution{PACTE, CNRS, Sciences Po Grenoble}
  \country{France}
}

\author{Olivier Bodini}
\affiliation{%
 \institution{LIPN, USPN}
 \country{France}
 }

\author{Manuel Moussallam}
\affiliation{%
  \institution{Deezer Research}
  \country{France}
}

\renewcommand{\shortauthors}{Matrosova, et al.}

\begin{abstract}
This paper examines the influence of recommender systems on local music representation, discussing prior findings from an empirical study on the LFM-2b public dataset \footnote{Previously available at \url{http://www.cp.jku.at/datasets/LFM-2b/}, the LFM-2b dataset has recently been taken down due to license issues.}. This prior study argued that different recommender systems exhibit algorithmic biases shifting music consumption either towards or against local content.
However, LFM-2b users do not reflect the diverse audience of music streaming services.
To assess the robustness of this study's conclusions, we conduct a comparative analysis using proprietary listening data from a global music streaming service, which we publicly release alongside this paper. We observe significant differences in local music consumption patterns between our dataset and LFM-2b, suggesting that caution should be exercised when drawing conclusions on local music based solely on LFM-2b.
Moreover, we show that the algorithmic biases exhibited in the original work vary in our dataset, and that several unexplored model parameters can significantly influence these biases and affect the study's conclusion on both datasets. Finally, we discuss the complexity of accurately labeling local music, emphasizing the risk of misleading conclusions due to unreliable, biased, or incomplete labels. To encourage further research and ensure reproducibility, we have publicly shared our dataset and code.

\end{abstract}




\maketitle

\section{Introduction}
Recommender systems are essential for music streaming services like Apple Music, Deezer, and Spotify~\cite{schedl2018current,jacobson2016music,schedl2021music,briand2021semi}. They help mitigate information overload problems by showcasing the most relevant content for each user, within large catalogs of millions of songs, albums, and artists~\cite{schedl2018current,li2023recent,bobadilla2013recommender,ferraro2023murs,hansen_recsys20}. They also assist users in discovering new music they might like on these services~\cite{briand2024let,bendada2023track,ferraro2023murs}. With the rise of streaming as the predominant form of music consumption~\cite{hiller2017rise,ifpi2023}, there has been, however, a noticeable increase in debates about the responsibilities of these systems. Concerns are also growing about their ability to promote a fair and diverse musical landscape and the various biases they might introduce or amplify when recommending music~\cite{lesota2021analyzing, ferraro2019music,kowald2020unfairness,bauer2017music,ferraro2019artist,shakespeare2020exploring,dinnissen2022improving,dinnissen2022fairness,ab2}.

In particular, Lesota et al.~\cite{lesota2022traces} recently argued that some music recommender systems might intensify the predominance of US music consumption in other countries. 
Specifically, in an empirical study focused on the LFM-2b dataset of listening actions on Last.fm~\cite{schedl2022lfm}, these authors investigated the extent to which standard recommender systems favor US-produced content over \textit{local} music from the country of origin of each user. 
Their findings suggest that NeuMF~\cite{he2017neural}, a neural network-based collaborative filtering algorithm, recommends lower proportions of local music than what users from each country actually listen to. In other words, NeuMF exhibits an \textit{algorithmic bias}~\cite{ab1,ab2} against local music on LFM-2b. On the contrary, the more classical ItemKNN~\cite{deshpande2004item} method yields more calibrated recommendations, and even fosters the consumption of local music in most countries under consideration.

This study undoubtedly raised essential issues regarding the uneven impact of recommender systems on local music consumption. Nonetheless, as big as it might be, the LFM-2b dataset used for evaluation is of a particular nature. Last.fm users tend to be active on the Internet and social media, and are not evenly distributed across countries. Therefore, they might not represent the full spectrum of music streaming service users.
Moreover, as detailed in Section~\ref{s2}, the study did not analyze how important parameters related to model training impact these biases. For these reasons,
it remains unclear whether the conclusions of Lesota et al.~\cite{lesota2022traces} would hold in other experimental settings and using a different dataset.

In this paper, we address this question by conducting a comprehensive comparative study using proprietary listening data  from the global music streaming service Deezer. Our study concentrates on France, Germany, and Brazil -- three countries where Deezer is one of the leading market players. Our contributions are as follows:
\begin{itemize}
    \item Firstly, we show that the Deezer and LFM-2b datasets present different local music consumption patterns. This discrepancy suggests caution when drawing conclusions about local music representation based solely on one dataset~like~LFM-2b.
    \item Secondly, we demonstrate that NeuMF and ItemKNN exhibit different algorithmic biases towards local music on Deezer compared to LFM-2b when following the evaluation setup of Lesota et al.~\cite{lesota2022traces}. Importantly, we also uncover several factors that significantly influence these biases -- including their magnitude but also their direction -- thereby affecting this study's overall conclusions on both datasets. These factors include the number of tracks each model recommends, their training variability, and whether they were trained on data from individual countries or the entire dataset.
    
    \item Thirdly, we explain that accurately labeling local music is a complex endeavor, and that the proportion of local music consumed and recommended can vary significantly depending on the source of the labels, their level of completeness, and the various biases introduced by human annotators.
    Consequently, we recommend prioritizing the development of comprehensive, transparent, and reliable local data labeling in future research. We argue that this foundational step is crucial for studies aiming to understand local music biases, as results based on unreliable labels may be misleading.
    
    \item Lastly, along with this paper, we publicly release our Deezer dataset as well as the source code of our experiments. This release aims to ensure full reproducibility of our results and to facilitate future studies on local music recommendation.
\end{itemize}

The remainder of this paper is organized as follows. In Section~\ref{s2}, we introduce the problem more formally and review the related work in more detail. In Section~\ref{s3}, we introduce our Deezer dataset and compare it to LFM-2b in terms of local music consumption. We report and discuss results from our empirical study on local music recommendation and biases in Section~\ref{s4}, and conclude in Section~\ref{s5}.

\section{Preliminaries}
\label{s2}

We begin this section by formally presenting the problem under consideration, before reviewing the related work.

\subsection{Problem Formulation}
\label{s21}

\subsubsection{Notation}
In this paper, we consider a set $\mathcal{V}$ of music tracks available in the catalog of a music streaming service, and a set $\mathcal{U}$ of $M \in \mathbb{N}^*$ users on this same service. 
We denote by $N_{\text{listened}}(u)$ the number of streams performed by each user $u$ over a predefined time period. 
Moreover, we denote by $N_{\text{local}}(u) \in \{0,\dots,N_{\text{listened}}(u)\}$ the number of these streams that are of music tracks from the country of origin of $u$ according to some data labeling\footnote{We note that associating music tracks with specific countries can be an ambiguous task, and that different data labeling rules may yield inconsistent results. The importance of the labeling source will be  pointed out and further discussed throughout this paper.}. We refer to $N_{\text{local}}(u)$ as the number of \textit{local} streams of $u$. 
Using this formalism, the proportion of local music listened to by $u$ is:
\begin{equation}
\text{L}(u) = \frac{N_{\text{local}}(u)}{N_{\text{listened}}(u)} \in [0, 1].
\end{equation}
Additionally, we consider a music recommender system: 
\begin{equation}
\text{MRS}_K \colon \mathcal{U} \to \mathcal{V}.
\end{equation}
$\text{MRS}_K$ recommends\footnote{At this stage, we do not formulate assumptions regarding the specific data or paradigm (e.g., a collaborative filtering or content-based approach~\cite{bobadilla2013recommender}) used to recommend tracks.} $K$ music tracks from $\mathcal{V}$ to each user of the music streaming service, for some fixed value $K < N$. 
The number of local music tracks recommended to the user $u$ by $\text{MRS}_K$ among these $K$ tracks is $N_{\text{local,MRS}_K}(u) \in \{0,\dots, K\}$. The proportion of local music tracks recommended to $u$ by $\text{MRS}_K$ is:
\begin{equation}
\text{L}_{\text{MRS}_K}(u) = \frac{N_{\text{local,MRS}_K}(u)}{K} \in [0, 1].
\end{equation}

\subsubsection{Objective}
Our main goal in this paper is to investigate the impact of $\text{MRS}_K$ on local music representation. In line with Lesota et al.~\cite{lesota2022traces}, our main indicator of interest will be the \textit{algorithmic bias} of $\text{MRS}_K$ in favor or against local music, defined as follows:
\begin{equation}
\text{Bias}_{\text{MRS}_K}  = \frac{1}{M} \sum_{u \in \mathcal{U}} \Big( \text{L}_{\text{MRS}_K}(u) - \text{L}(u) \Big),
\label{eq-bias}
\end{equation}
with $\text{Bias}_{\text{MRS}_K} \in [-1, 1]$. In essence, a positive bias (respectively, a negative bias) indicates that, on average, $\text{MRS}_K$ recommends more local music (resp., less local music) than what users of the music streaming service organically listen to. The remainder of this paper will analyze this value for different datasets, recommender systems, and settings. We will aim to uncover the various factors that might influence the intensity or even the direction~of~this~bias. 

\subsection{The study of Lesota et al.~\cite{lesota2022traces} on LFM-2b}
\label{s22}

Analyzing local music algorithmic biases was one of the key objectives of Lesota et al.~\cite{lesota2022traces}, with a particular emphasis on the predominance of US music consumption in other countries.

\subsubsection{Context}
 Over the past decades, US music has dominated the global music industry, with its cultural influence spreading worldwide. Trends from the US have been widely adopted even in local music productions \cite{fuhr2015globalization}, and the ratio of US music on radio charts has been rapidly increasing\footnote{However, this growth slowed down from the 1990s due to several factors, including the emergence of CDs, which made music production more accessible worldwide, content localization by MTV, and the introduction of laws in countries like France, imposing local music quotas on radio station programming~\cite{achterberg2011cultural}.} since the 1960s \cite{achterberg2011cultural}.
The consequences of this dominance are mixed~\cite{achterberg2011cultural,crane2014cultural,castells2011power}. On the one hand, it can potentially stimulate local cultural development through the adaptation to global trends and reinforcement of local identity, a process known as \textit{glocalization}~\cite{achterberg2011cultural,castells2011power}. On the other hand, it is also sometimes perceived as a threat, termed \textit{cultural imperialism}, which could lead to the decline of local cultures~\cite{crane2014cultural,morley2006globalisation,tomlinson2001cultural}.
While music has become more centralized with the rise of music streaming services~\cite{bello2021cultural,lesota2022traces}, at the time of Lesota et al.'s study~\cite{lesota2022traces}, limited research had focused specifically on the impact of these services and their recommender systems on local music consumption.

\subsubsection{Results}
In 2022, Lesota et al.~\cite{lesota2022traces} published results from their empirical study conducted on a subset\footnote{The entire LFM-2b dataset includes approximately 2 billion listening events over 15 years from about 120~000 users. Lesota et al.~\cite{lesota2022traces} analyzed a subset of 14 million interactions from 2018 and 2019, involving 13~000 users in 20 countries selected for having at least 100 users and artists who had collectively created at least 1~000 tracks.} of the LFM-2b public dataset, which includes listening events from users of the Last.fm music website~\cite{schedl2022lfm}.
This study explored the prominence of US cultural imperialism in online music consumption, revealing that while the US maintains a strong position among Last.fm users, its influence varies significantly across countries. The authors also observed varying glocalization patterns depending on countries. The final part of their study, which our reproducibility paper focuses on, investigated whether recommender systems can increase existing predominances and, overall, shift music consumption towards specific countries at the expense of local content.
The authors answered this question positively, explaining that the influence is uneven and algorithm-dependent. Their experiments suggest that NeuMF~\cite{he2017neural}, a neural network-based collaborative filtering algorithm, recommends lower proportions of local music than what Last.fm users in each country organically listen to. In other words, NeuMF exhibits a negative $\text{Bias}_{\text{MRS}_K}$ local bias, as computed in Equation~\eqref{eq-bias}. Conversely, the more traditional ItemKNN~\cite{deshpande2004item} method tends to promote the consumption of local music in most  countries, i.e., it is associated with a positive $\text{Bias}{\text{MRS}_K}$ local music bias.

\subsubsection{Limitations of Lesota et al.~\cite{lesota2022traces} and Motivations of our Work} \label{limits}
This study undoubtedly raises important issues regarding the uneven impact of recommender systems on local music representation. It positions itself within a growing body of scientific research focusing on the fairness of music recommender systems and their biases, not only regarding local music but also other aspects, including gender and music genres~\cite{lesota2021analyzing, ferraro2019music,shakespeare2020exploring,ferraro2019artist,dinnissen2022fairness,ab2}.

Nonetheless, we believe this study also suffers from limitations that motivate our work.
From an algorithmic perspective, the authors investigated biases using a single number of recommended tracks, $K=10$. The effect of varying $K$, i.e., allowing their systems to recommend different numbers of tracks, on the magnitude or even the direction of biases is unclear. Additionally, systems like the neural network-based NeuMF include randomness in training~\cite{he2017neural}, yet the robustness of the study's conclusions to this randomness remains unverified. Thirdly, their systems were trained on all users, but using data solely from users of each specific country might alter the findings. Bauer and Schedl~\cite{bauer2018importance, bauer2019global}, for instance, suggest distinguishing between country-specific and global mainstreamness to ensure a realistic representation of musical preferences in different countries and promote less biased recommendations.

Beyond these algorithmic considerations, replicating this study with a different dataset is also worthwhile. Indeed, Last.fm users tend to be active on the internet and social media and are not evenly distributed across countries~\cite{schedl2022lfm}. Therefore, they might not reflect the diverse audience of music streaming
services.
Furthermore, Lesota et al.~\cite{lesota2022traces} relied on MusicBrainz, an open music encyclopedia~\cite{musicbrainz}, to associate artists with country labels. However, MusicBrainz labels may not only be imprecise but are also missing for some LFM-2b tracks, which were simply excluded by Lesota et al.~\cite{lesota2022traces}. Zanger et al.~\citep{zanger2023risk} suggest that this exclusion could lead to measurement errors due to label biases. Indeed, less popular artists or those from locations or genres unrepresented among MusicBrainz human annotators might lack more labels\footnote{As an illustration, Jul, a French rapper, has no music genre annotation in MusicBrainz, despite being one of the most-streamed artists on Deezer in France during 2018 and 2019.}, leading to a distorted representation of local music in some countries. These factors raise the question of whether the findings of Lesota et al.~\cite{lesota2022traces} would remain valid with actual listening data from a music streaming service, or by using alternative labeling sources. The remainder of this paper will aim to clarify these aspects.




\section{Comparing LFM-2b To Deezer Data}
\label{s3}

In this section, we present our dataset of listening events from Deezer and subsequently provide a comparative analysis with~LFM-2b. 

\subsection{The Deezer Dataset}

\subsubsection{Overview} 
\label{s311} We examine a proprietary dataset from the global music streaming service Deezer, comprising the listening history of 30~000 randomly selected users on this platform in March 2019. The dataset features an equal distribution of users from the three countries of our study -- France, Germany, and Brazil -- with 10~000 users from each. It includes approximately 4 million streams across over 565~000 distinct music tracks. To maintain consistency with Lesota et al.~\cite{lesota2022traces}, we did not filter listening events by streaming context. As a result, the dataset includes both organic streams and recommendations, which we later discuss in Section~\ref{s4}.

\subsubsection{Local Music Labeling} The country of each user is determined based on their IP address.
Determining an artist's country may sometimes be ambiguous, for example in the case of artists born in a country but who gained fame in another one. To capture this complexity, we consider three different country labels in our work:
\begin{itemize}
    \item Our dataset includes the main \textit{country of activity} and \textit{country of origin} of each artist, as provided by Deezer when available, and compiled by this service from public and private sources.
    \item Moreover, we added the publicly accessible country labels from the MusicBrainz open music encyclopedia~\cite{musicbrainz} when available. We recall that these labels were the ones used by Lesota et al.~\cite{lesota2022traces} in their original study on LFM-2b.
\end{itemize}

\begin{figure}[t]
\centering
    \includegraphics[width=0.85\columnwidth]{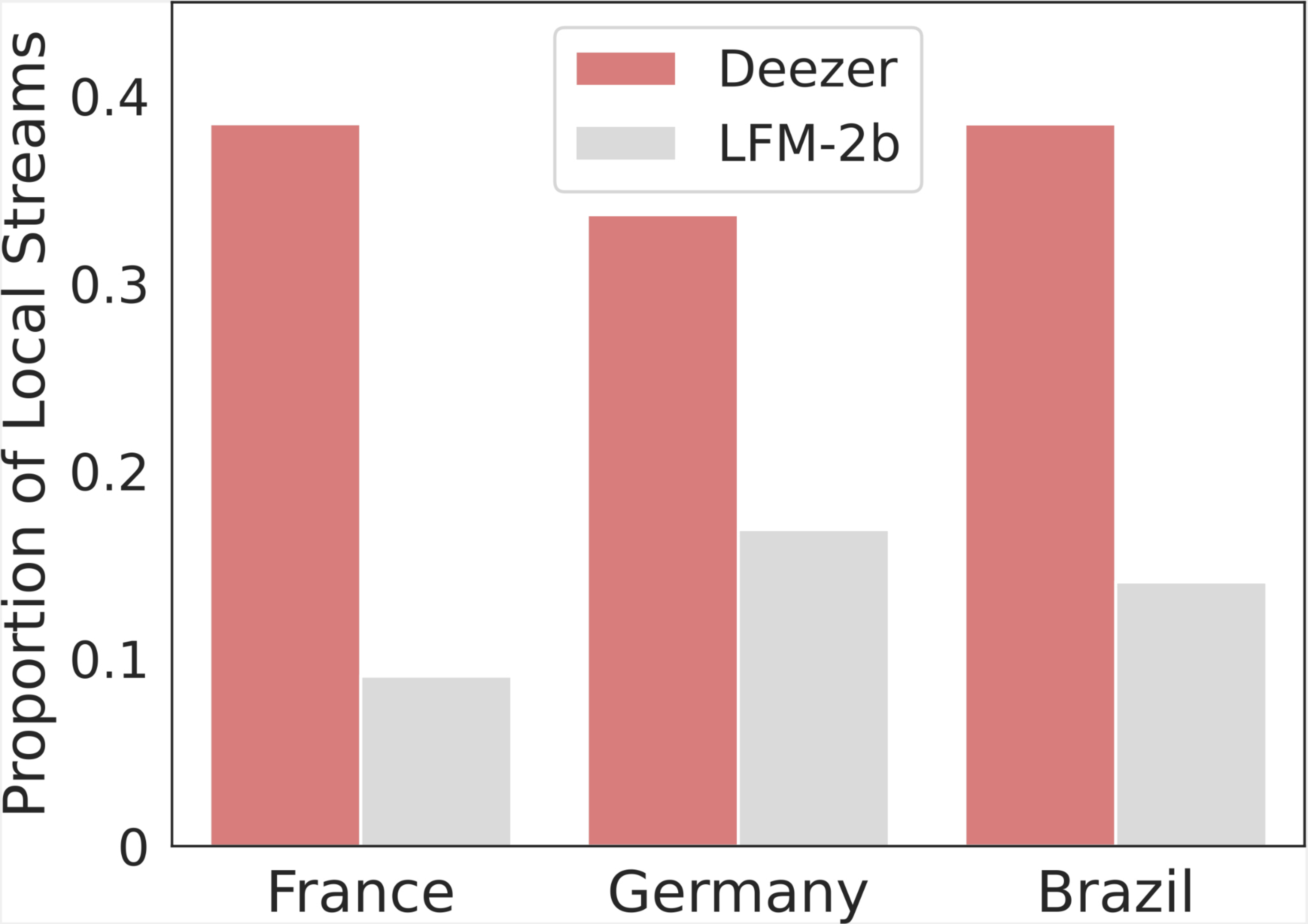}
    \caption{Proportion of local streams by country, according to the LFM-2b and Deezer datasets. All values are computed using MusicBrainz labels, by considering labeled tracks only.}
    \label{fig1}
\end{figure}

\begin{table}[t]
\caption{Top 10 most streamed music tracks by French users, in the LFM-2b (top) and Deezer (bottom) datasets. All reported country labels originate from MusicBrainz.}
\begin{center}
 \resizebox{\columnwidth}{!}{
\begin{tabular}{c|c|c|c|c|c|c} 
\toprule
\textbf{Dataset} & \textbf{Artist/Band} & \textbf{Title} & \makecell{\textbf{Country}\\ \textbf{Label}} & \makecell{\textbf{Singing}\\ \textbf{Language}} & \makecell{\textbf{Release} \\ \textbf{Year}} & \makecell{\textbf{Music} \\ \textbf{Genre}} \\
\midrule
\midrule
\multirow{10}{*}{\textbf{LFM-2b}}& Portishead & Glory Box & GB & EN & 1994 & Trip Hop\\
& Radiohead & Karma Police & GB & EN & 1997 & Alt. Rock\\
& The Verve & Bitter Sweet Sym. & GB & EN & 1997 & Alt. Rock\\
& Franz Ferdinand & Take Me Out & GB & EN & 2004 & Indie Rock\\
& a-ha & Take On Me & NL & EN & 1985 & Synth Pop\\
& Angèle & Balance ton quoi & BE & \textbf{FR} & 2018 &  Pop\\
& The xx & Intro & GB & EN & 2009 & Trip Hop\\
& 4 Non Blondes & What's Up? & US & EN & 1993 & Alt. Rock\\
& Metronomy & The Look & GB & EN & 2010 & Indie Rock\\
& Wax Tailor & Que Sera & \textbf{FR} & EN & 2004 & Trip Hop\\
\midrule
\midrule
\multirow{10}{*}{\textbf{Deezer}} & Ninho & Goutte d'eau & \textbf{FR} & \textbf{FR} & 2019 &  Rap\\ 
& Angèle & Tout oublier & BE & \textbf{FR} & 2018 &  Pop\\
& Lady Gaga & Shallow & US & EN & 2018 & Folk Pop\\
& Lomepal & Trop beau & \textbf{FR} & \textbf{FR} & 2018 &  Rap/Pop\\
& David Guetta & Say My Name & \textbf{FR} & EN & 2018 & EDM\\
& Ariana Grande & 7 rings & US & EN & 2019 & Pop\\
& Alonzo & Assurance vie & \textbf{FR} & \textbf{FR} & 2019 &  Rap\\
& DJ Snake & Taki Taki & \textbf{FR} & EN & 2018 & EDM\\
& Kaaris & Gun salute & \textbf{FR} & \textbf{FR} & 2019 &  Rap\\
& Booba & PGP & \textbf{FR} & \textbf{FR} & 2019 & Rap\\
\bottomrule
\end{tabular}
}
\end{center}
\label{tab1}
\end{table}

\begin{figure*}[t]
    \centering
     \subfigure[LFM-2b - France - MusicBrainz]{\includegraphics[scale=0.3]{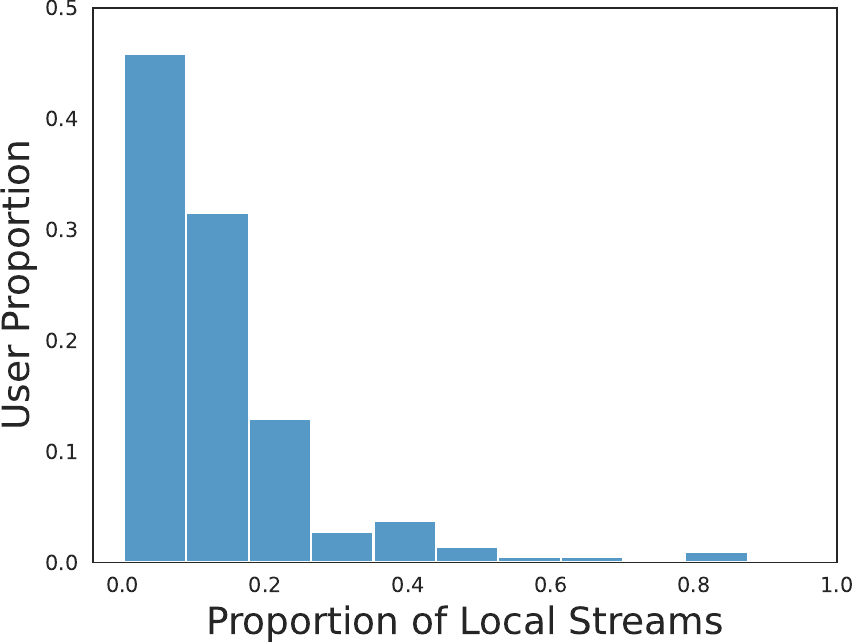}}
    \subfigure[Deezer - France - MusicBrainz]{\includegraphics[scale=0.3]{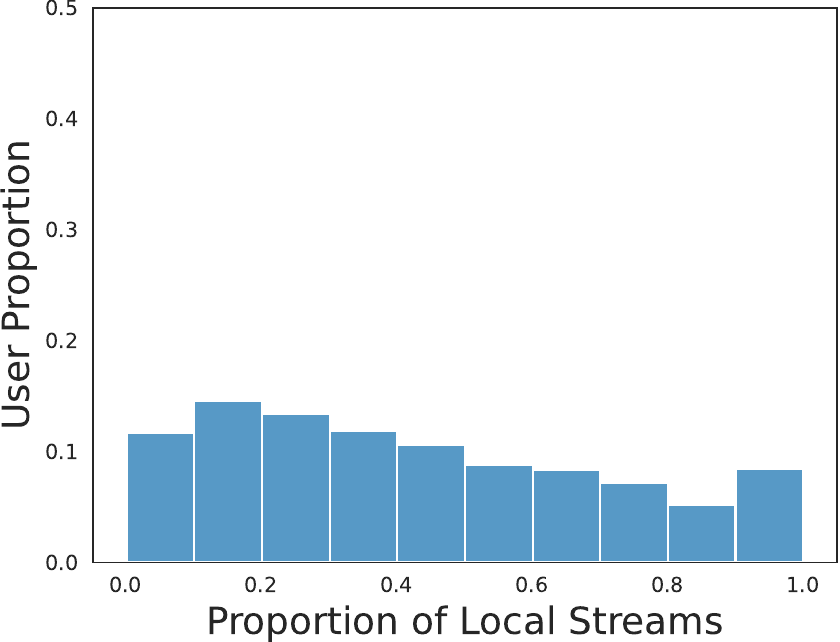}}
    \subfigure[Deezer - France - Active]{\includegraphics[scale=0.3]{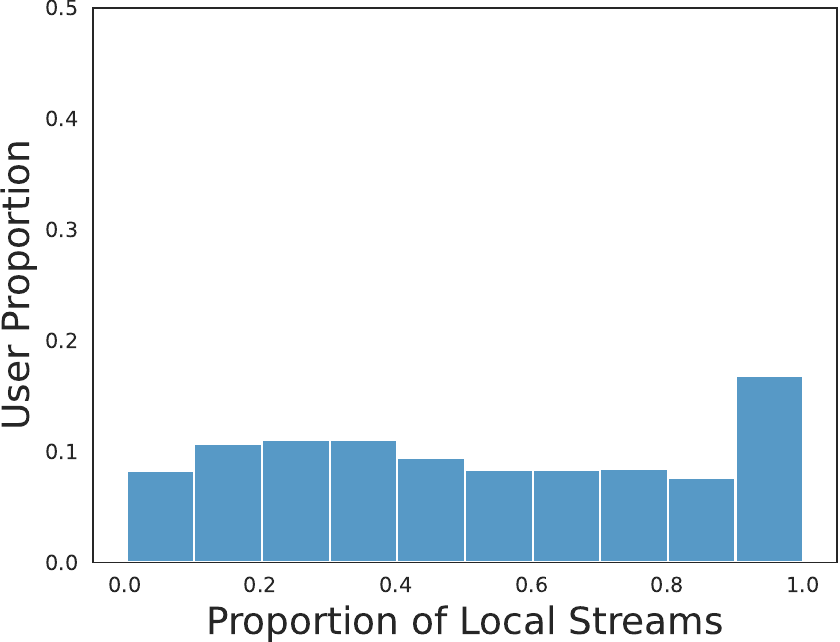}}
    \subfigure[Deezer - France - Origin]{\includegraphics[scale=0.3]{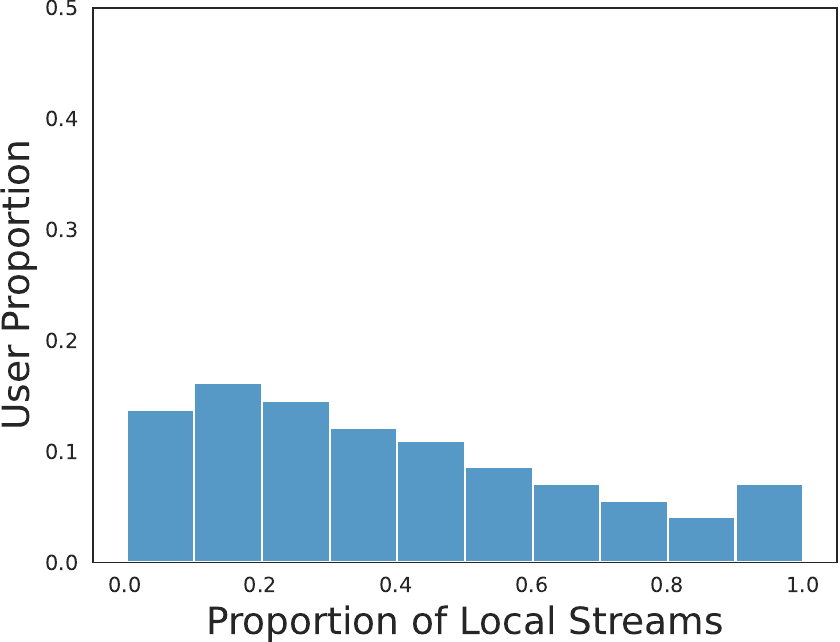}}

    \subfigure[LFM-2b - Germany - MusicBrainz]{\includegraphics[scale=0.3]{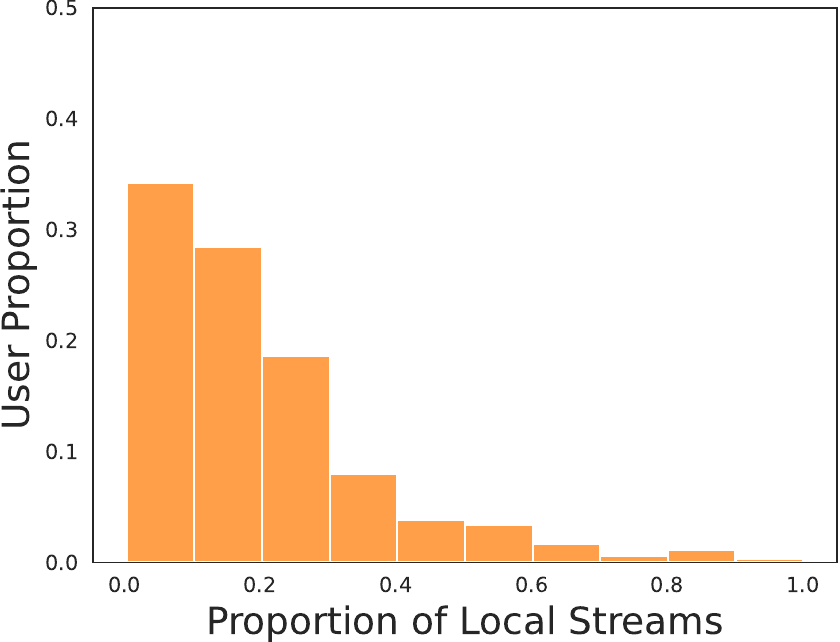}}
    \subfigure[Deezer - Germany - MusicBrainz]{\includegraphics[scale=0.3]{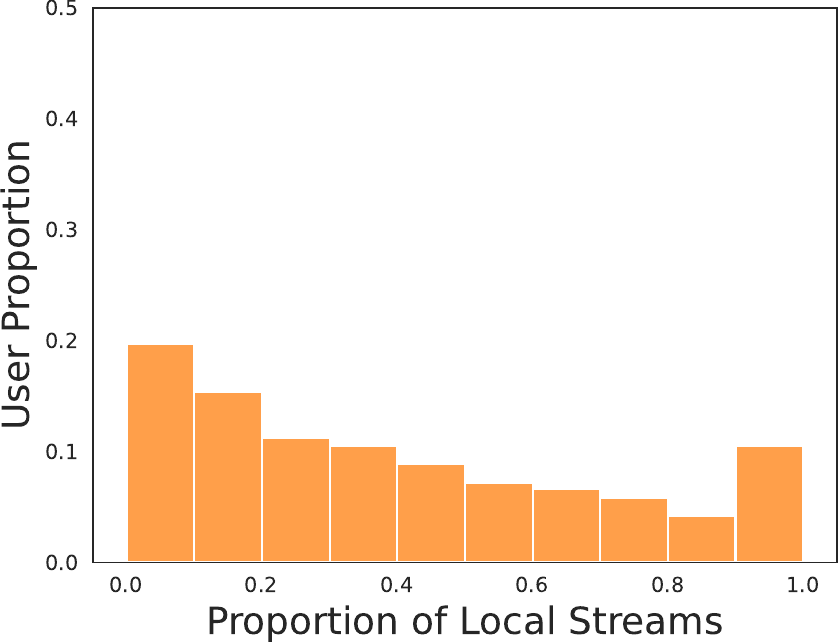}}
    \subfigure[Deezer - Germany - Active]{\includegraphics[scale=0.3]{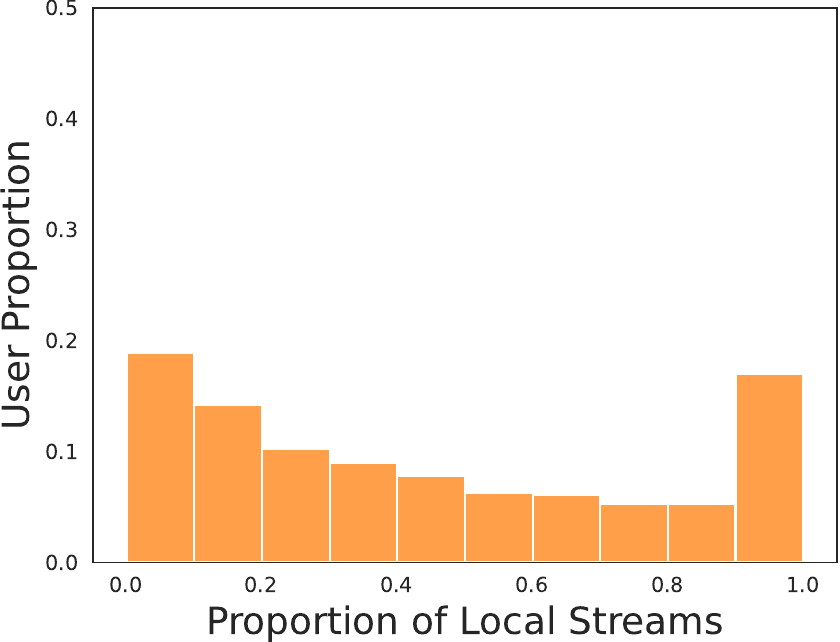}}
    \subfigure[Deezer - Germany - Origin]{\includegraphics[scale=0.3]{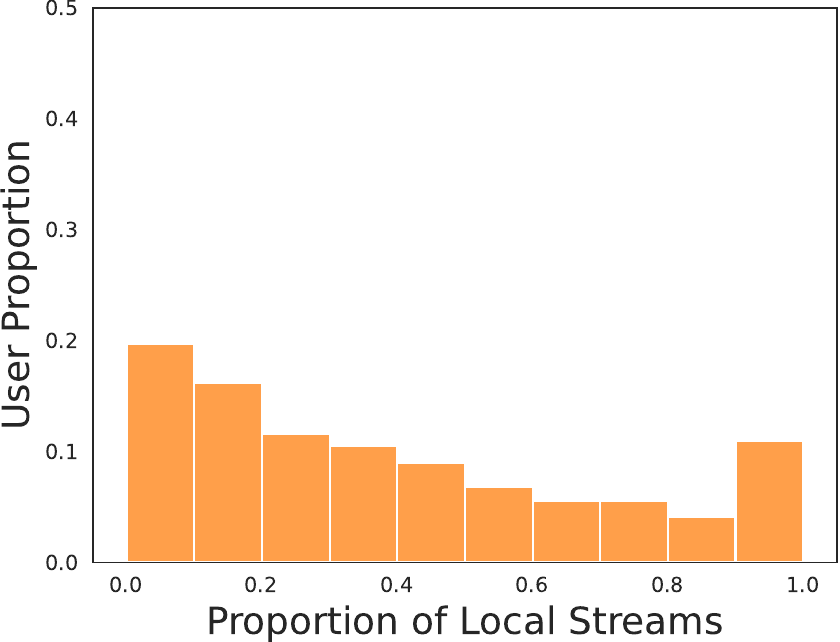}}

    \subfigure[LFM-2b - Brazil - MusicBrainz]{\includegraphics[scale=0.3]{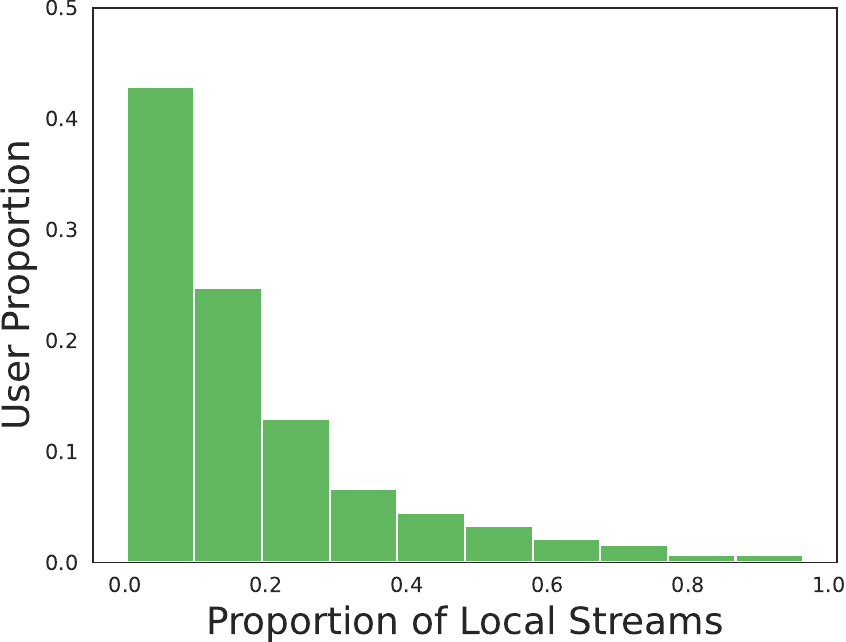}}
    \subfigure[Deezer - Brazil - MusicBrainz]{\includegraphics[scale=0.3]{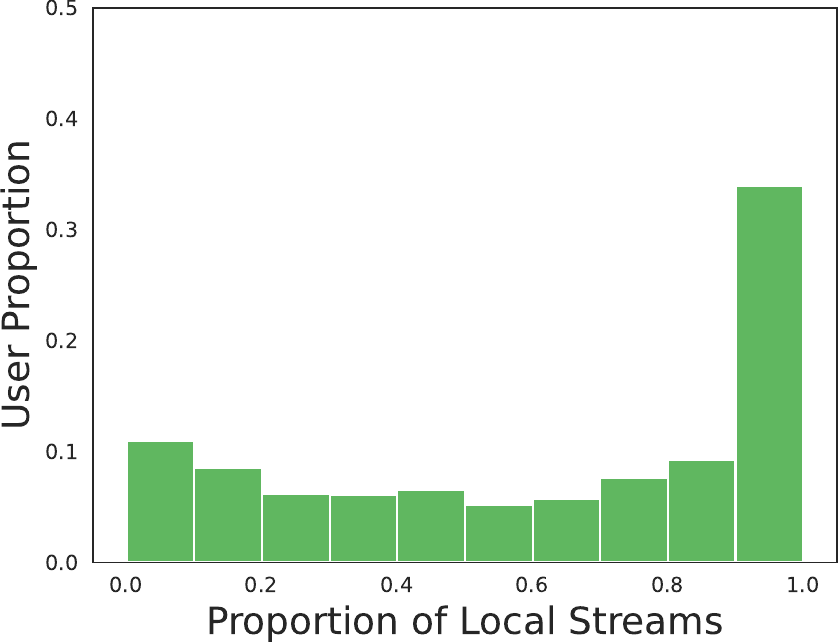}}
    \subfigure[Deezer - Brazil - Active]{\includegraphics[scale=0.3]{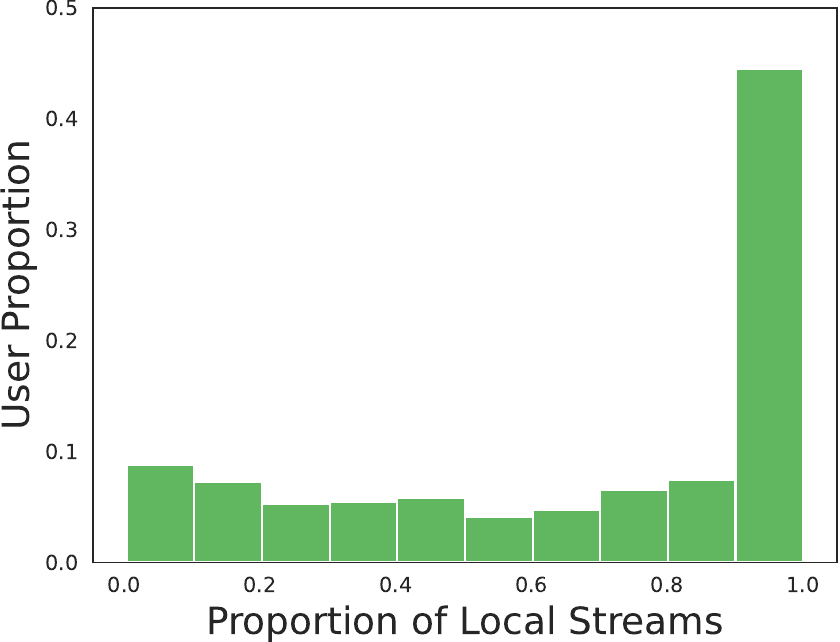}}
    \subfigure[Deezer - Brazil - Origin]{\includegraphics[scale=0.3]{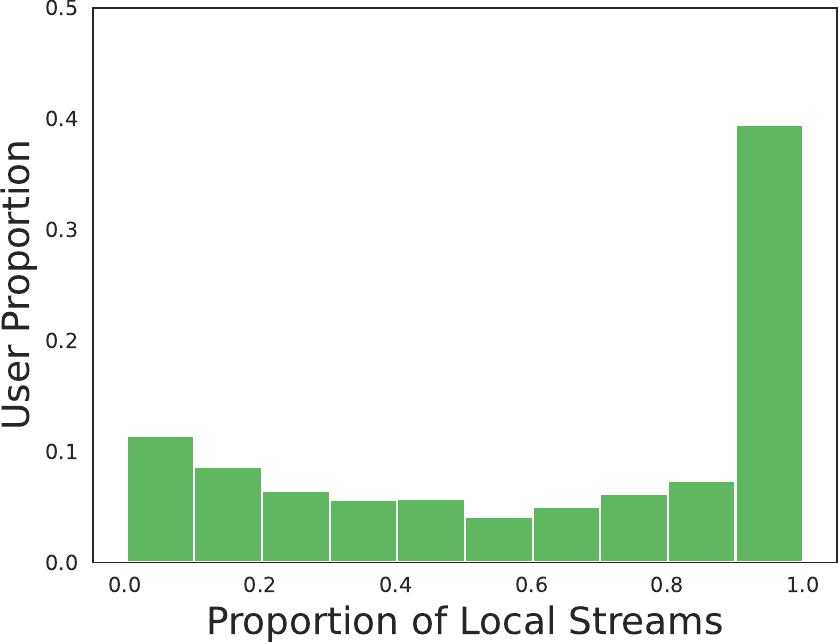}}
    \caption{Histograms of the proportion of local streams per user (considering labeled tracks only). Results are split by dataset (i.e., LFM-2b or Deezer), country (i.e., France, Germany, or Brazil), and labeling source (i.e., MusicBrainz labels, Deezer's country of activity, or Deezer's country of origin).}
    \label{fig2}
\end{figure*}

\subsection{Descriptive Analysis}

We now provide a descriptive analysis of our Deezer dataset. We compare it to the subset of LFM-2b\footnote{We use data kindly provided by the authors via private email communications.} of Lesota et al.~\cite{lesota2022traces} and additionally restrict this subset to the three countries of interest in this study. The sample contains {254} users from France, {805} users from Germany, and {1064} users from Brazil, for a total of over 3 million listening events in 2018 and 2019, on around 100~000 music tracks.

\subsubsection{LFM-2b vs Deezer}
Overall, we observe that Deezer and LFM-2b exhibit quite different patterns of local music consumption.

First of all, by plotting the proportion of local streams in both
datasets, using MusicBrainz labels and considering only labeled
tracks in both cases (Figure~\ref{fig1}), we observe a significantly lower rate
of local music in the LFM-2b dataset. For Brazil, for instance, LFM-2b
exhibits 2.5 times fewer local streams compared to Deezer. Moreover,
we report in the first two columns of Figure~\ref{fig2} histograms of the percentage of local streams per user, here again using MusicBrainz labels. We note
that the distributions are different between the two datasets. In the Deezer
dataset, across all three countries, users exhibit varying patterns:
some do not listen to local music at all, while others listen to
local music only, with a large spectrum of behaviors in between. Conversely,
the LFM-2b dataset shows a stark contrast, with few users
listening to a majority of local music.

To go further, we present in Table~\ref{tab1} the top 10 most streamed music tracks in France, in both datasets. 
The Deezer dataset not only contains a higher proportion of French music (both in terms of artists and lyrics), but also predominantly features recent releases, from the year prior to, or the same year as the streams. The prevalent genres include pop, rap, and electronic music. On the contrary, the top tracks in the LFM-2b dataset predominantly consist of older releases (dating back one or more decades) with English lyrics and genres such as indie, alternative rock, or trip hop, which are more niche.
These differences can be attributed to several characteristics of the Last.fm website, upon which the LFM-2b dataset is built~\cite{schedl2022lfm}. Firstly, Last.fm caters primarily to music enthusiasts who have a strong inclination towards collecting and organizing their music libraries, potentially resulting in a preference for less mainstream music genres. Secondly, this website's users are predominantly English-speaking persons, which introduces a population bias.
While the Deezer dataset may appear to be more reflective of realistic music consumption patterns, it is important to acknowledge the possibility of similar biases existing within it, as well as in data from other streaming services. Hence, it is crucial to proceed with caution when asserting the presence of cultural patterns based on such data. Using multiple data sources for cross-validation becomes imperative to ensure the reliability and accuracy of conclusions.

\begin{table}[t]
\caption{Percentages of (i) labeled streams, (ii) local streams (among the labeled streams) and (iii) local streams (among all streams) in the Deezer dataset, by country and label source. A labeled stream corresponds to a stream of a music track associated with a country label. A local stream corresponds to a stream where the user and the artist have the same country label.}
\begin{center}
\resizebox{\columnwidth}{!}{%
\begin{tabular}{c|c|c|c|c} 
\toprule
\textbf{Country} & \textbf{Label Source} & \makecell{\textbf{Labeled}\\\textbf{Streams}} & \makecell{\textbf{Local Streams}\\\textbf{Among Labeled}} & \makecell{\textbf{Local Streams}\\\textbf{Among All}}\\
\midrule
\midrule
\multirow{3}{*}{France} 
& Deezer - Activity	& 76 \% &  50 \% &  38 \%\\
& Deezer - Origin   & 75 \% &  34 \% &  26 \%\\ 
& MusicBrainz       & 76 \% &  38 \% &  29 \%\\
\midrule
\multirow{3}{*}{Germany} 
& Deezer - Activity & 60 \% &  40 \% &  24 \%\\
& Deezer - Origin	& 62 \% &  30 \% &  18 \%\\
& MusicBrainz       & 69 \% &  33 \% &  23 \%\\
\midrule
\multirow{3}{*}{Brazil} 
& Deezer - Activity & 41 \% &  48 \% &  19 \%\\
& Deezer - Origin   & 36 \% &  37 \% &  13 \%\\
& MusicBrainz       & 38 \% &  38 \% &  14 \%\\
\bottomrule
\end{tabular}
}
\end{center}
\label{table2}
\end{table}

\subsubsection{Impact of Label Sources}
\label{s322}

Table~\ref{table2} presents the proportions of labeled streams and local streams (among the labeled ones, and among all streams) in the Deezer dataset, according to the three label sources, i.e., Deezer's country of origin and country of activity, as well as MusicBrainz labels.
We observe that none of the labeling sources provides complete coverage. Across the three countries considered, between 64\% and 24\% of the streams remain unlabeled. Streams in different countries exhibit varying levels of label coverage. For example, the artist's country is identified in 75-76\% of streams by French users, depending on the label source, while for streams from Brazil, this coverage drops to only 36-41\%. Label coverage varies by country depending on the source. For instance, Deezer's activity labels provide the highest coverage for streams from Brazil, but they offer the least coverage for streams from Germany.
Furthermore, the proportions of local consumption strongly vary depending on the label source. For instance, considering only labeled streams, only 38\% of French users' streams consist of French tracks according to MusicBrainz labels, whereas Deezer's activity labels indicate 50\%. This difference is notable given that both sources have identical label coverage rates for French streams. 
Due to incomplete labeling, there's a significant difference between the proportions of local streams among labeled streams versus all streams. Calculating local streams based solely on labeled data can suggest higher local consumption than what is actually observed across all streams. Moreover, these values aren't proportional; for instance, when considering only labeled streams, France and Brazil vie for the title of the largest local consumer (depending on the label source). However, when all streams are taken into account, Brazil exhibits the lowest local consumption by a substantial margin.

In summary, obtaining a complete and universally unquestionable labeling of local music proves to be a challenging task. Overall, we advocate against simply filtering out unlabeled tracks, as done in the reference study~\cite{lesota2022traces}. Indeed, such an approach may result in removing a majority of the streams from the study, potentially undermining the validity of the study's conclusion. As outlined in Section~\ref{limits}, this filtering operation can also introduce label biases~\cite{zanger2023risk}, when annotators exhibit preferences for specific countries, music genres, or languages during the annotation process. The extent to which these discrepancies between labels result in inconsistencies regarding the measurement of local music algorithmic biases will be analyzed in Section~\ref{s4}.

\section{Experimental Analysis of Local Music Recommendation and Biases}
\label{s4}

In this section, we now present our empirical analysis of local music recommendation and biases on the LFM-2b and Deezer datasets. We start by describing the experimental setting. Then, we report and discuss our findings. Notably, we compare them with the main conclusions of the original study of Lesota et al.~\cite{lesota2022traces}.

\subsection{Experimental Setting}
\label{s41}

\subsubsection{Models}

We examine the same two collaborative filtering~\cite{koren2015advances} recommender systems analyzed in the reference study:
\begin{itemize}
    \item NeuMF~\cite{he2017neural} is a deep learning-based recommender system that integrates traditional matrix factorization (MF) \cite{koren2015advances} with neural networks \cite{goodfellow2016deep}. NeuMF first learns \textit{embedding} vector representations of users and music tracks in the same vector space where proximity reflects similarity, by factorizing a user-track interaction matrix. These embeddings are then processed by a deep neural network architecture trained to predict the most relevant tracks to recommend to each user.
    \item ItemKNN~\cite{deshpande2004item} is a more traditional recommender system based on the nearest neighbor approach~\cite{koren2015advances}. For each user, it assigns similarity scores to tracks by evaluating how similar these tracks are to those the user has previously listened to. This similarity is determined in a collaborative filtering fashion, by analyzing the interactions of other users. When recommending a set of $K$ tracks, ItemKNN selects the top $K$ neighbors with the highest similarity scores. Unlike NeuMF, ItemKNN operates directly on the user-track interaction matrix and does not learn embedding representations.
\end{itemize}

Lesota et al.~\cite{lesota2022traces} trained their ItemKNN and NeuMF models using the entire LFM-2b dataset, including users from all countries. In contrast, our study not only trains these models on the complete LFM-2b and Deezer datasets but also considers country-specific variants. We developed these variants by using only listening data from users within the same country -- a realistic setting for a global music streaming service. For instance, to recommend music to a Deezer French user, we would employ the ItemKNN or NeuMF variant trained on Deezer's listening data from French users.

\begin{figure*}[th]
    \centering
    \subfigure[LFM-2b - Global - MusicBrainz]{\includegraphics[scale=0.265]{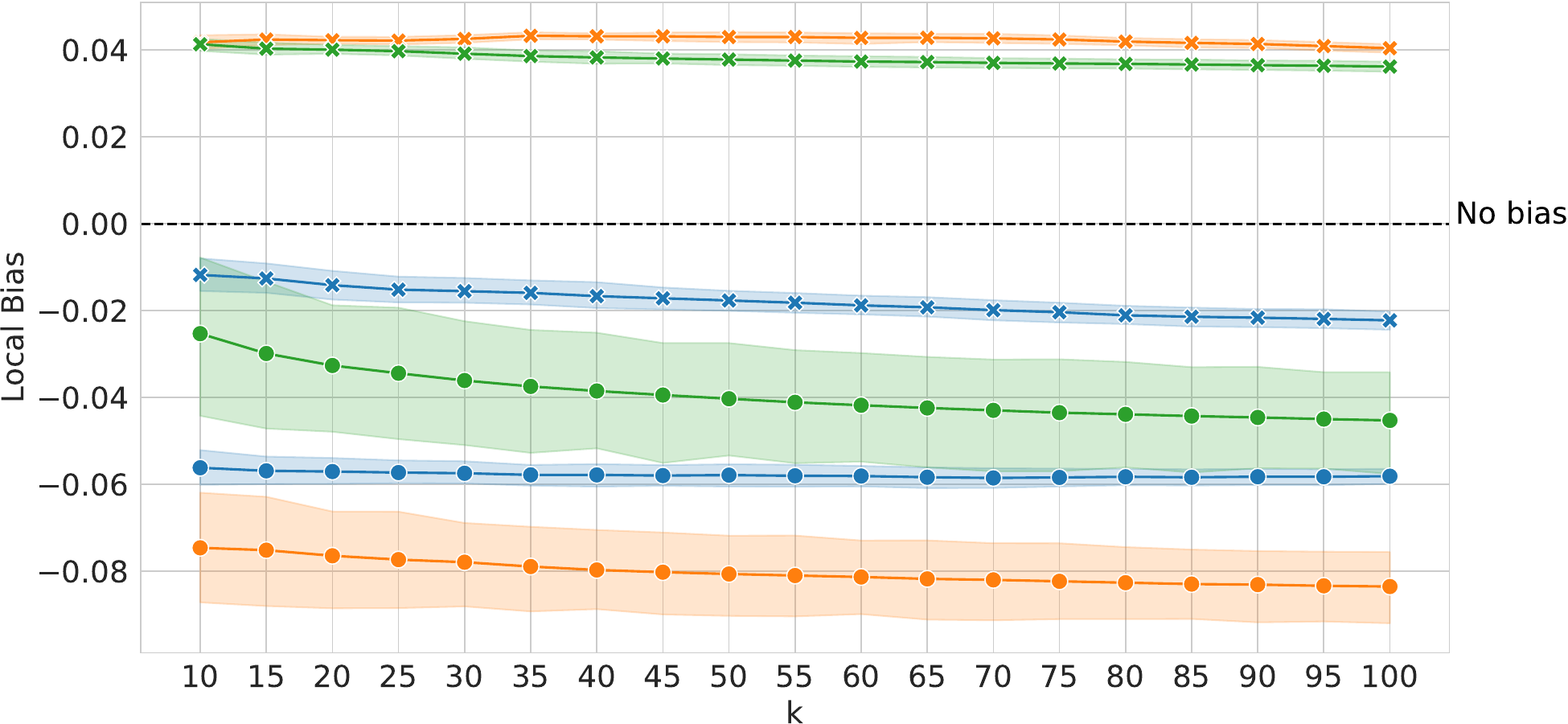}}
    \subfigure[LFM-2b - Local - MusicBrainz]{\includegraphics[scale=0.265]{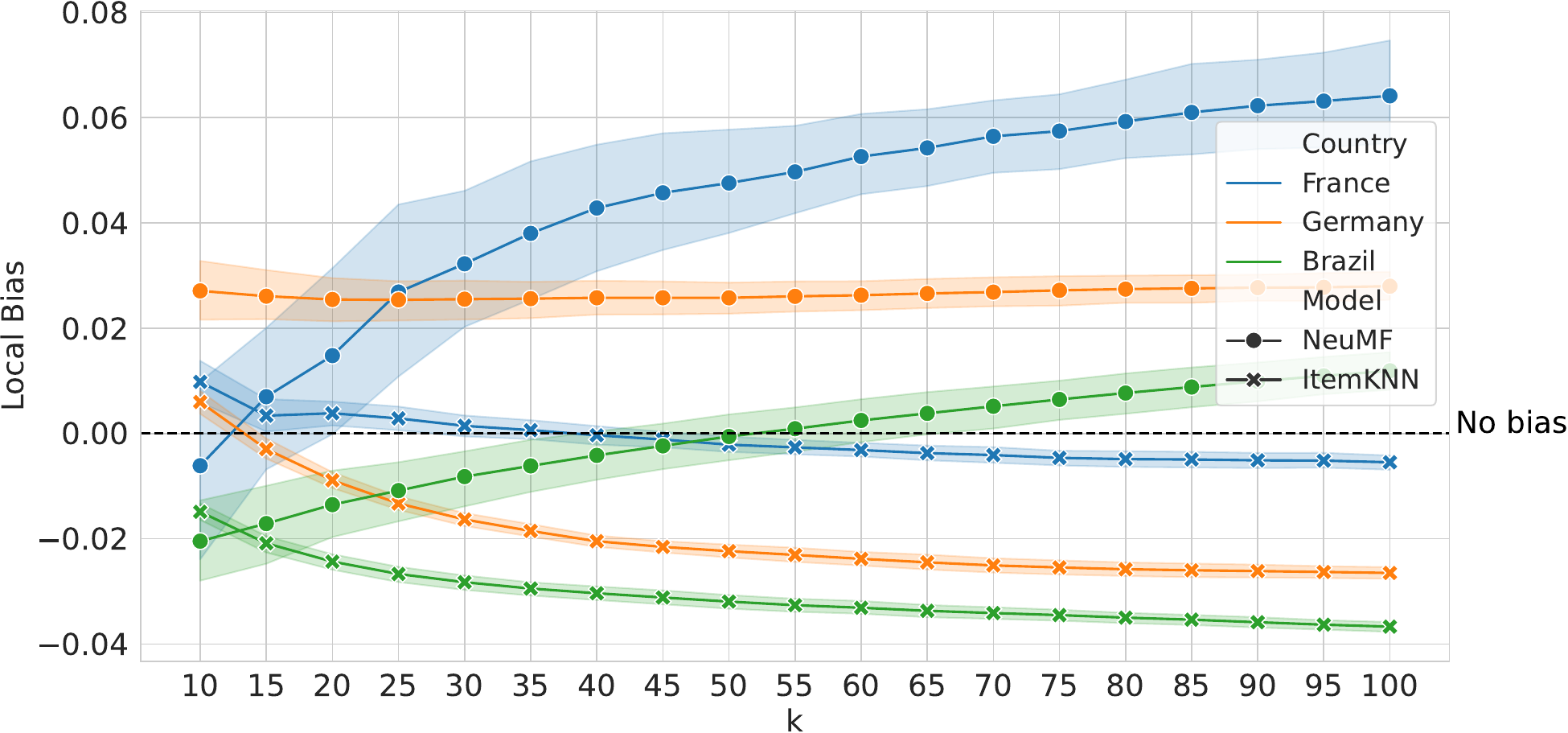}}
    \caption{Local music algorithmic biases of ItemKNN and NeuMF on LFM-2b users in France, Germany, and Brazil, computed for numbers of recommended tracks $K$ varying from 10 to 100 with a step of 5 tracks. Results are split by training variant (``Global'' models are trained using listening data from users of all countries, while ``Local'' models are trained using only listening data from users of the same country). All values are averaged over 20 model runs and reported with $\pm$ 1 standard deviation intervals. Values above (respectively, under) the ``No bias''  0-level horizontal dotted line indicate that the model exhibits a positive (resp., a negative) algorithmic bias towards local music.}
    \label{fig3}
\end{figure*}

\subsubsection{Task and Implementation Details} 
For both datasets, we train all models on a top 10 track recommendation task, evaluated using mean reciprocal rank (MRR@10) scores~\cite{zangerle2022evaluating} computed on a validation set of 10\% randomly selected users, masked during the training phase. Afterwards, we compute the local biases $\text{Bias}_{\text{MRS}_K}$, as defined in Equation~\eqref{eq-bias}, for each model in each country, averaged across all users in that country\footnote{We note that one might alternatively compute biases using only users from a test set. However, reporting biases for all users not only aligns with the evaluation protocol of Lesota et al.~\cite{lesota2022traces}, but also reflects the practical goal of music streaming services, which would typically aim to address local biases across their entire user base.}. While in the reference study~\cite{lesota2022traces}, the authors only reported results for a single number of recommended tracks ($K=10$), here we consider the more general case of a varying $K$, with $K$ ranging from 10 to 100 with a step~of~five~tracks.

As in the reference study, we use the implementation of ItemKNN and NeuMF available in RecBole~\cite{zhao2021recbole}, a Python library based on PyTorch~\cite{paszke2019pytorch} that aims to provide a unified framework for developing and reproducing recommendation algorithms. For ItemKNN, we retrieve nearest neighbors to recommend by using cosine similarities computed from the user-track train interaction matrix, with a null value for the shrink parameter~\cite{shrinkref}. We train all NeuMF models for a maximum of 300 epochs using the Adam optimizer~\cite{kingma_iclr15}, with a learning rate of 0.001, batch sizes of 512 items, a dropout rate of 0.1~\cite{srivastava2014dropout}, and minimizing a binary cross-entropy loss~\cite{he2017neural}. All NeuMF models learn embedding vectors of dimension 64. For interested readers, we provide exhaustive information on each layer of every neural network in our public GitHub repository (see Section~\ref{implem}).

\subsection{Results and Discussion}
\label{s42}

\subsubsection{Results on LFM-2b}
\label{resultslfm}

We begin our analysis with the results obtained on the LFM-2b public dataset.
We report in Figure~\ref{fig3} the local music algorithmic biases of ItemKNN and NeuMF on LFM-2b with MusicBrainz labels, averaged over 20 model runs with standard deviations to assess variability in the training process.
Results are split by training variant, i.e., global or country-specific.
In particular, Figure~\ref{fig3}(a) reports results for the global ItemKNN and NeuMF variants trained on users from all countries, which matches the specific setting of Lesota et al.~\cite{lesota2022traces} (with $K=10$ only in their study).
Overall, we reproduce results comparable to those of the original study in this specific setting. NeuMF recommends lower proportions of local music than what users from France, Germany, and Brazil listen to, unveiling negative algorithmic biases. In contrast, ItemKNN tends to foster the consumption of local music in Brazil and Germany, while displaying a negative but relatively small bias in France. Our results are consistent when modifying the number~$K$ of recommended tracks.

However, Figure~\ref{fig3}(b) reveals that the results change drastically when training ItemKNN and NeuMF in a country-specific fashion, i.e., using data from LFM-2b but selecting users of a single country only, instead of all users. For instance, for $K = 10$, NeuMF now shows a positive bias in Germany, while ItemKNN shows a negative bias in Brazil. Interestingly, increasing the number of recommended tracks $K$ can also reverse the bias direction. For example, NeuMF exhibits a negative bias for $K \in \{10, \dots, 45\}$ but a positive bias for $K \in \{50, \dots, 100\}$. This observation highlights the importance of testing different values of $K$ to draw robust conclusions about the potential biases of each recommender system against local music. Lastly, Figure~\ref{fig3}(b) underlines the importance of accounting for variability in the training process, particularly for NeuMF, which shows large $\pm 1$ standard deviation intervals (this variability primarily stems from randomness in the initialization of neural weights, dropout components, and the use of different training splits for each model run). As an illustration, in France and for $K = 10$, NeuMF's interval overlaps with the ``No bias'' horizontal dotted line. This emphasizes that NeuMF has shown both positive and negative biases in our experiments, depending on each training instance.

\begin{figure*}[ht]
    \centering
    \subfigure[Deezer - Global - MusicBrainz]{\includegraphics[scale=0.265]{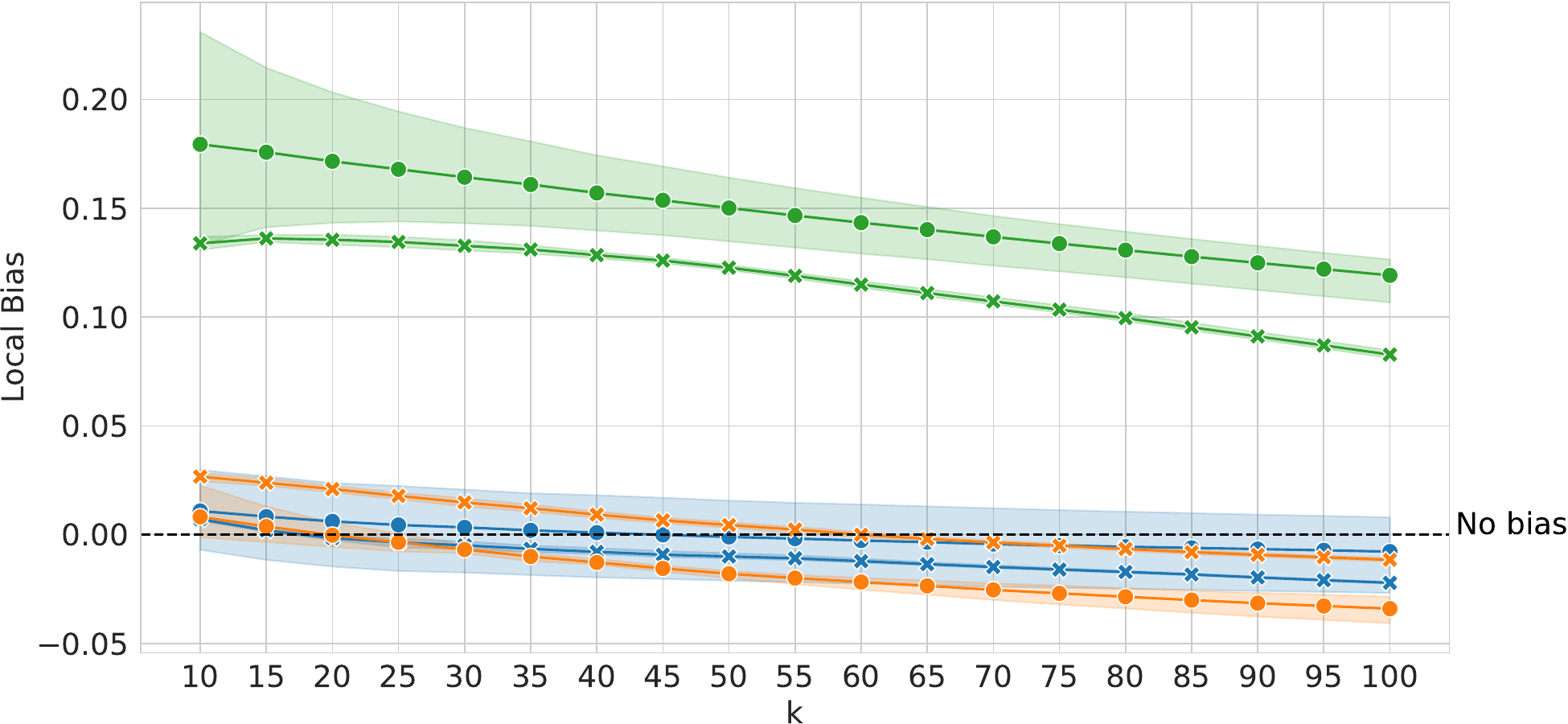}}
    \subfigure[Deezer- Local - MusicBrainz]{\includegraphics[scale=0.265]{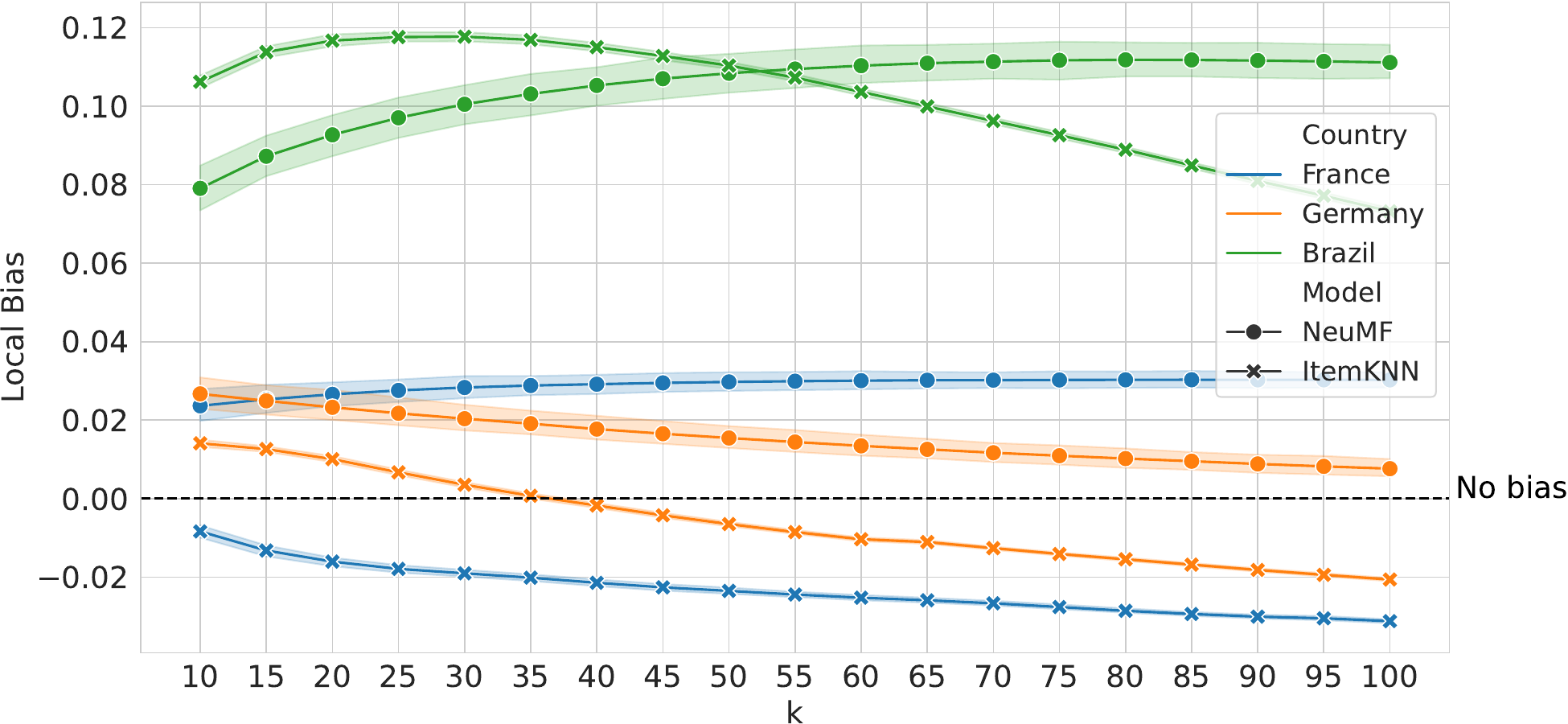}}
    
    \subfigure[Deezer - Global - Activity]{\includegraphics[scale=0.265]{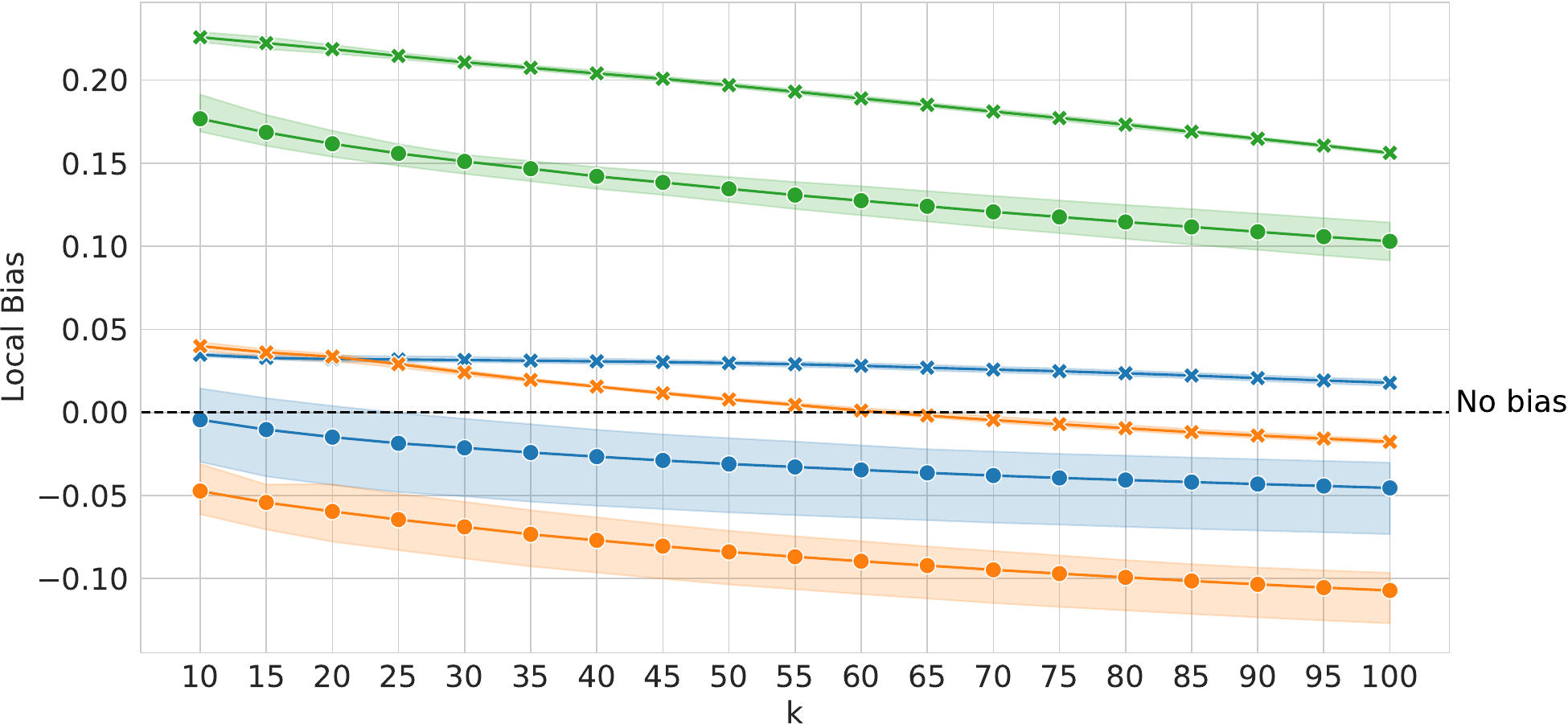}}
    \subfigure[Deezer - Local - Activity]{\includegraphics[scale=0.265]{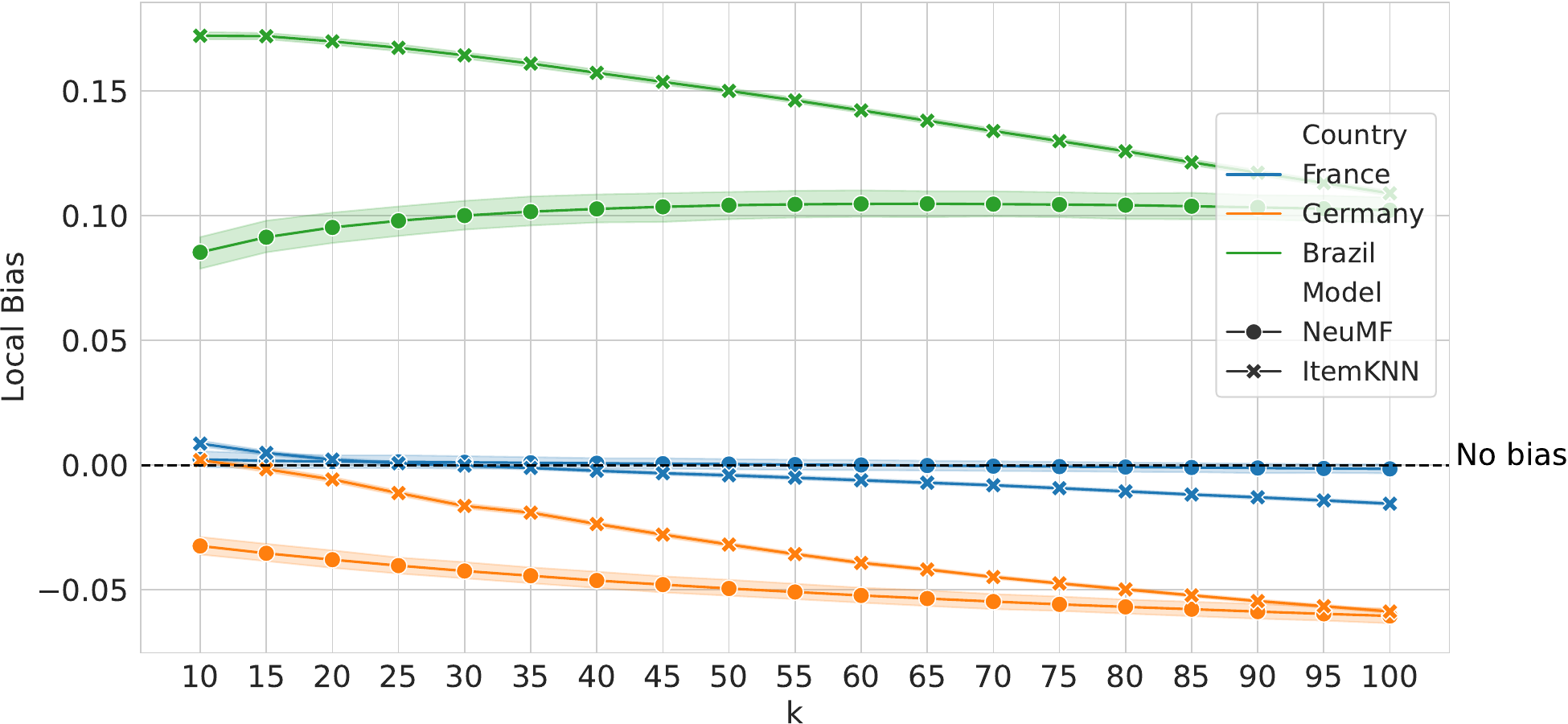}}
    \subfigure[Deezer - Global - Origin]{\includegraphics[scale=0.265]{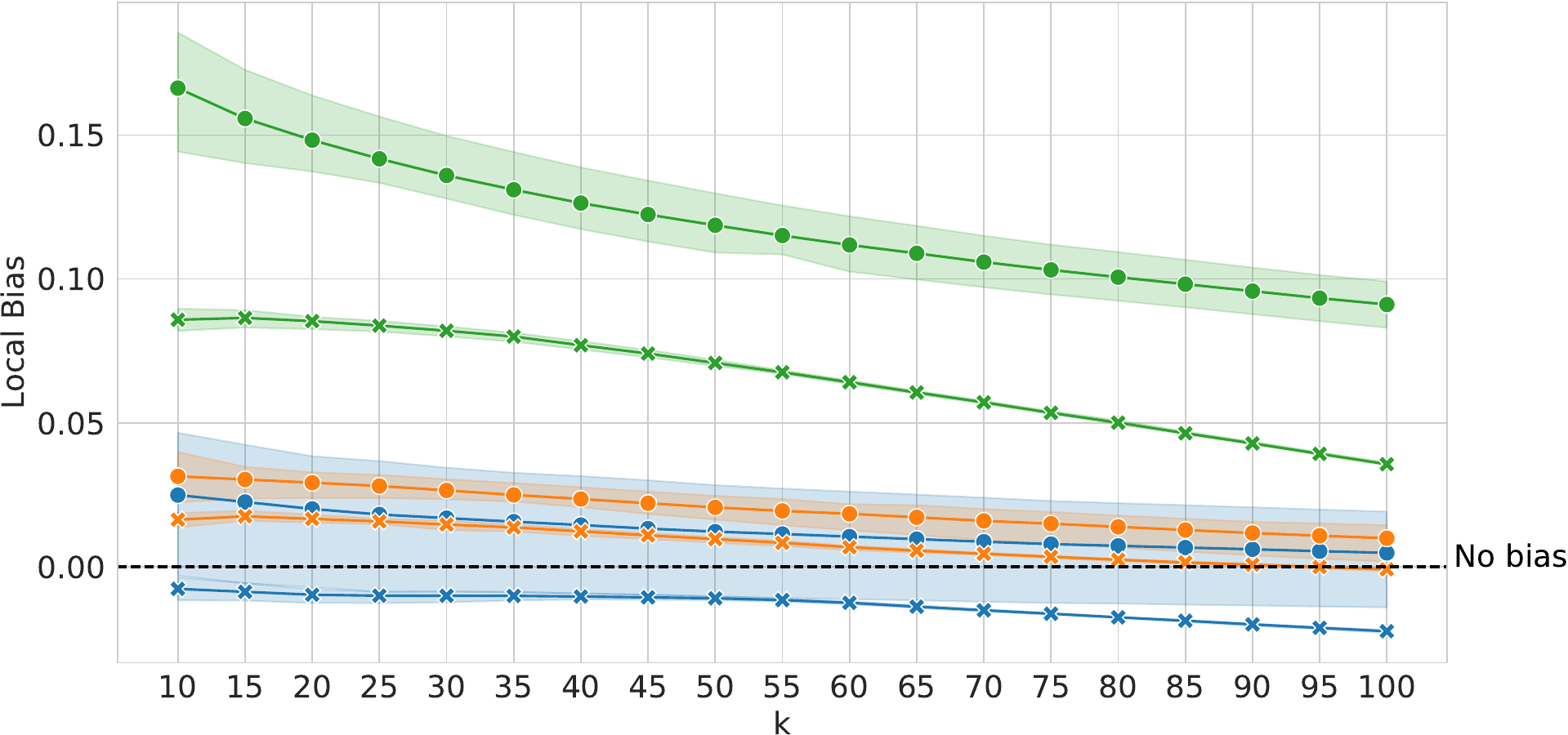}}
    \subfigure[Deezer - Local - Origin]{\includegraphics[scale=0.265]{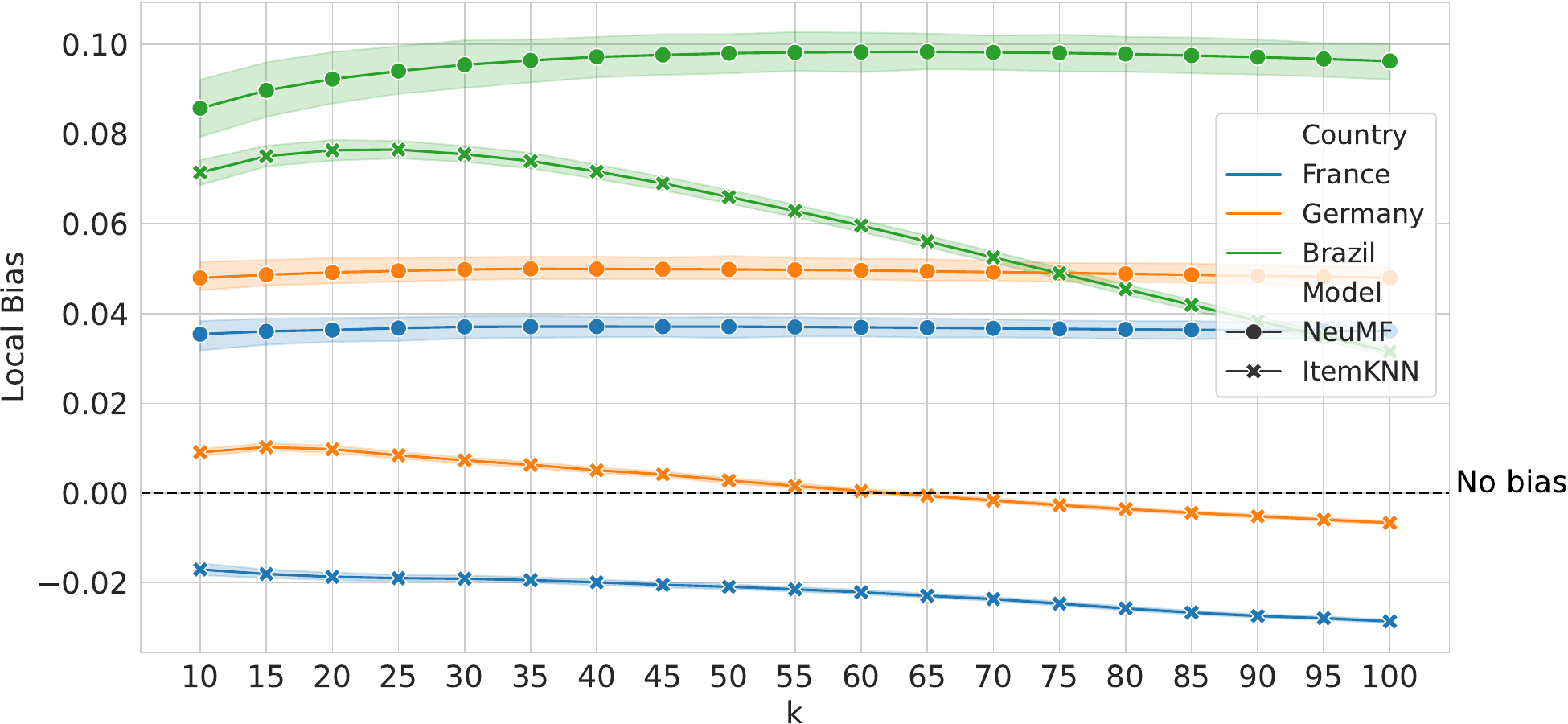}}
    
    \caption{Local music algorithmic biases of ItemKNN and NeuMF on Deezer users in France, Germany, and Brazil, computed for numbers of recommended tracks $K$ varying from 10 to 100 with a step of 5 tracks. Results are split by training variant (``Global'' models are trained against listening data from users of all countries, while ``Local'' models are trained using only listening data from users of the same country), and by label source (i.e., MusicBrainz labels, Deezer's country of activity, or Deezer's country of origin). All values are averaged over 20 model runs and reported with $\pm$ 1 standard deviation intervals. Values above (resp. under) the ``No bias'' 0-level dotted line indicate that the model exhibits a positive (resp. negative) algorithmic bias.}
    \label{fig4}
\end{figure*}

\subsubsection{Results on Deezer}
We now compare these results with those obtained using the proprietary Deezer dataset, which contains streams from users of the music streaming service Deezer.
Figure~\ref{fig4} presents the local music algorithmic biases of ItemKNN and NeuMF on this dataset. Once again, all biases are averaged over 20 model runs with standard deviations. Results are categorized by training variant (i.e., global or country-specific) and label source (i.e., MusicBrainz labels, Deezer's country of activity, or Deezer's country of origin).
We begin our discussion with an inspection of Figure~\ref{fig4}(a), which displays results for the global ItemKNN and NeuMF variants using MusicBrainz labels. This setting is consistent with the one used in the reference study~\cite{lesota2022traces} and our earlier Figure~\ref{fig3}(a), but applied to the Deezer dataset instead of LFM-2b.
We observe that the algorithmic biases exhibited by ItemKNN and NeuMF on LFM-2B vary significantly on Deezer. At $K = 10$, for instance, all models are associated with a positive average bias value, contrary to previous results from Figure~\ref{fig3}(a) and Lesota et al.~\cite{lesota2022traces} on LFM-2b. We also notice that biases tend to be of higher magnitude on Deezer users. These apparent discrepancies tend to reinforce our discussion from Section~\ref{s3}, highlighting the need for caution when drawing conclusions based solely on one dataset like LFM-2b. While this public dataset serves as a convenient starting point for researchers, the observed biases may not be consistent across different datasets with varying listening patterns.

Figure~\ref{fig4}(b) reports comparable experiments using our local variants of ItemKNN and NeuMF with MusicBrainz labels. Below, Figure~\ref{fig4}(c) and Figure~\ref{fig4}(d) present results for global and local models, respectively, but using Deezer's country of activity labels instead of MusicBrainz labels. Finally, Figure~\ref{fig4}(e) and Figure~\ref{fig4}(f) show results for global and local models, respectively, but using Deezer's country of origin. Overall, these figures confirm our previous insights from Section~\ref{resultslfm}. Training models with data coming from one country only can significantly alter local music biases in both magnitude and direction. Additionally, changing the number of recommended music tracks $K$ and using different user splits or weight initializations can also affect these biases.
While these algorithmic factors were not examined in the original work~\cite{lesota2022traces}, our experiments reveal that they can change the global picture and the study's conclusion. Properly accounting for these factors is, therefore, crucial to ensure a robust and reliable analysis of local music algorithmic biases.

Figure~\ref{fig4} 
also highlights that changing the label sources can also substantially affect conclusions. Section~\ref{s3} had already uncovered that the proportions of local music consumed by users may strongly depend on the label source. Figure~\ref{fig4} further demonstrates that this variation leads to inconsistencies in the measurement of local music algorithmic biases. For instance, the global and local NeuMF models are consistently associated with a negative bias in Germany when relying upon the (country of) ``Activity'' label (Figure~\ref{fig4}(c) and Figure~\ref{fig4}(d)), but on the opposite they are associated with a positive bias when using the (country of) ``Origin'' label (Figure~\ref{fig4}(e) and Figure~\ref{fig4}(f)).
While changing the label source drastically impacts the results, we acknowledge that, nonetheless, some findings remain robust to these changes. For example, Brazil is consistently associated with the highest positive biases across almost all settings in Figure~\ref{fig4}. We hypothesize that this consistent behavior may be due to the higher number of Brazilian Deezer users who listen exclusively or almost exclusively to local music,  according to all three label sources (see Figure~\ref{fig2}). This aspect might be reflected in our models, although further analysis would be required for confirmation.

\subsubsection{Limitations and Future Work}
\label{s42n}
In concluding our discussion, we acknowledge some limitations of our experimental analysis, which also offer opportunities for future research. Firstly, as explained in Section~\ref{s311}, our Deezer dataset includes both organic streams and recommended streams, to maintain consistency with the original study. However, focusing solely on organic streams for model training could be worthwhile. As Lesota et al.~\cite{lesota2022traces} have also pointed out, incorporating recommended tracks may distort the model's insights about user preferences. Examining the impact of this adjustment on biases towards local music would be an interesting avenue for further investigation.
Secondly, Villermet et al.~\cite{villermet2021follow} showed that users are very different when it comes to their use of the different features offered by streaming platforms -- including algorithmic recommendation -- and that only a minority of them primarily relies on algorithmic recommendation to select the music they listened to on streaming platforms. Thus, reproducing the experiment by considering only users who interact with music recommender systems to a certain degree could provide more relevant findings. Thirdly, while we used a specific definition of bias, its perception might actually be subjective, and individual user interviews may be useful to gain additional insights.

At first glance, one might also want to analyze the local music algorithmic biases of numerous other recommender systems, beyond ItemKNN and NeuMF. However, we believe that a more crucial preliminary step for future work will be to improve the accuracy of local music labels. Our study underscores the challenge of accurately labeling local music, and demonstrates how variations in label sources can substantially affect conclusions regarding local music representation and recommendation. Conducting extensive bias analyses on numerous recommender systems with the current state of labeling -- where no single label source covers more than 80\% of streams, and where labels reflect biases from human annotators~\cite{zanger2023risk} -- could prove fruitless. Indeed, results obtained from misleading labels could render such analyses unreliable. Consequently, we recommend prioritizing the development of comprehensive and reliable local data labeling in future research. We believe that cross-referencing assumptions across multiple labels remains one of the most reliable practices towards achieving this~goal.

\subsection{Open-Source Code and Data Release}
\label{implem}

Along with this paper we release two important resources. Firstly, an anonymized version of our Deezer proprietary dataset, containing 4~million listening events from 30~000 Deezer users in France, Germany, and Brazil, along with all three local music labels from our work\footnote{Dataset available at \url{https://zenodo.org/records/13309698}.}. The release of this industrial dataset aims to foster future research activities on music recommender systems and local music consumption analysis.
In addition, we are open-sourcing the entire Python source code of our experimental analysis, to ensure the reproducibility of our results. All materials are publicly available on GitHub \footnote{Code available at \url{https://github.com/kmatrosova/FairnessRecsys2024}}.

\section{Conclusion}
\label{s5}

In conclusion, although LFM-2b~\cite{schedl2022lfm} is publicly available and serves as a convenient starting point for researchers inspecting music recommender systems, caution should be exercised when drawing conclusions about local music consumption based solely on this dataset. Our paper has emphasized significant differences in local music consumption patterns between LFM-2b and a proprietary dataset comprising the listening history of French, German, and Brazilian users of the music streaming service~Deezer.

By replicating Lesota et al.~\cite{lesota2022traces}'s investigation of algorithmic biases in recommender systems for local music, we have also demonstrated that the two collaborative filtering models they analyzed, NeuMF~\cite{he2017neural} and ItemKNN~\cite{deshpande2004item}, display varying biases on Deezer compared to LFM-2b. Moreover, we have identified several factors related to model training that had not been examined in this previous work and can significantly influence these biases, thereby modifying the study's overall conclusions.

Importantly, we have also explained that the proportion of local music consumed and recommended can vary significantly depending on the label source under consideration, its level of completeness, and biases introduced by human annotators~\cite{zanger2023risk}. While obtaining complete and universally accepted local music labels proves to be challenging, we have nonetheless recommended to prioritize the research in this direction. We have argued that this foundational labeling step is crucial for studies aiming to understand local music biases, as results based on unreliable labels may be misleading.

Overall, our work highlights the importance of using multiple model settings and data sources for cross-validation, and to ensure robust conclusions regarding the biases of music recommender systems -- not only for local music but also potentially for other aspects such as gender and music genres. As a consequence, we have decided to publicly release our Deezer dataset along with this paper, including listening logs and labels from all three sources used in our experiments.
We hope that this release of industrial resources will foster further research. As discussed in Section~\ref{s42n}, our results come with certain limitations that open up interesting avenues for future analyses. Investigating these future directions would undoubtedly contribute to better measuring and enhancing the fairness of recommender systems on music streaming services.

\section*{Funding information}

This paper has been realized in the framework of the 'RECORDS' grant (ANR-2019-CE38-0013) funded by the ANR (French National Agency of Research).

\bibliographystyle{ACM-Reference-Format}
\bibliography{RECSYS24-86}

\end{document}